\documentclass{IEEEtran}
\usepackage{cite}
\usepackage{amsmath,amssymb,amsfonts}
\usepackage{algorithmic}
\usepackage{graphicx}
\usepackage{textcomp}
\def\BibTeX{{\rm B\kern-.05em{\sc i\kern-.025em b}\kern-.08em
    T\kern-.1667em\lower.7ex\hbox{E}\kern-.125emX}}
\begin{document}
\title{Planar near-field antenna measurements\\ 
with a uniform step larger than half-wavelength}




\author{R. Moretta, \IEEEmembership{Member, IEEE}, F. Pascariello, G. Petraglia, M. Feo, M.A. Maisto, \IEEEmembership{Member, IEEE}
	
\thanks{Manuscript received February 11, 2024.\\ This work was supported in part by the European Union and the Italian Ministry of University and Research funding through Programma Operativo Nazionale Ricerca e Innovazione 2019/2020 under Grant B26C18000080005. (\textit{Corresponding author: Raffaele Moretta.})}
\thanks{Raffaele Moretta was a postdoctoral researcher at University of Campania \textquotedblleft Luigi Vanvitelli", 81031 Aversa (NA), Italy. He is now an Antenna Engineer at MBDA Italy, 80070 Bacoli (NA),  Italy (e-mail: raffaele.moretta@mbda.it).}
\thanks{Fabio Pascariello, Giovanni Petraglia, and Maurizio Feo are with MBDA, 80070 Bacoli (NA),  Italy (e-mail: fabio.pascariello@mbda.it, giovanni.petraglia@mbda.it, maurizio.feo@mbda.it).}
\thanks{Maria Antonia Maisto is with Dipartimento di Ingeneria, Università della Campania \textquotedblleft Luigi Vanvitelli", 81031 Aversa (NA), Italy (e-mail: mariaantonia.maisto@unicampania.it).}}

\maketitle

\begin{abstract}
In this paper, a new sampling scheme of the near field radiated by a planar source is proposed and assessed. More in detail, the paper shows a uniform sampling criterion that allows representing the near field over a plane with a number of measurements lower than the classical half-wavelength sampling.

At first, a discretization strategy of the near field based on the warping method is recalled from the literature.
The latter requires to collect a non-redundant number of field measurements that are non-uniformly arranged over the observation domain.

Despite this, the warping sampling scheme works well only if the measurement plane does not overcome the source. When the observation domain is larger, it does not predict the exact positions of the field samples at the edges of the measurement plane; accordingly, in these regions it is not possible to recover the near field behavior by the collected samples.

To overcome this drawback, a spatially varying oversampling is exploited. The latter is chosen in such a way that the resulting sampling becomes uniform. Such choice also ensures a growth of the sampling rate only at the edges of the observation domain permitting the retrieval of the near field by its samples.

Finally, numerical simulations based on experimental data corroborate the effectiveness of the approach in recovering both the near and the far field. 
\end{abstract}


\section{Introduction}
\label{sec:introduction}

In the framework of antenna testing, the radiation pattern of the antenna under test (AUT) can be retrieved by far field or near field measurements.

Far field measurements are carried in free space
at a distance fulfilling the far-field condition depending on the frequency and the source size.
Since such distance may be large (especially for large sources), a high transmitting power could be necessary. Moreover, far field measurements may be negatively affected by bad weather conditions, obstacle reflections, ground reflections, and other signals.


All these drawbacks highlight the need to test antennas in an indoor environment (like an anechoic chamber) under fully controlled conditions. To achieve this goal, two different strategies can be adopted.

The first one concerns the use of a compact range to produce the far field at a short distance \cite{compact1}. In this case, the far field pattern is obtained by collimating the field radiated by the AUT with a large reflector or a lens.

The second strategy consists of measuring directly the near field \cite{near1}.
Then, once the near field data have been acquired, the radiation pattern of the AUT is recovered by exploiting a suitable near-field far-field transformation (NFFFT) \cite{NFFFT1,NFFFT2,NFFFT3,NFFFT4,NFFFT5,NFFFT6,NFFFT7,NFFFT8,NFFFT9,NFFFT10}.

However, since near field measurements can be acquired with a simpler measurement setup, the second strategy is mainly used in antenna testing \cite{compact_near_field}. 

In order to reconstruct the radiation pattern by near field data, a proper measurement scheme must be set. Over the years different measurement schemes of the near field have been developed for the planar \cite{planar1} \cite{planar2}, cylindrical \cite{cylindrical1} \cite{cylindrical2}, and spherical scanning \cite{spherical1, spherical2, spherical3, spherical4 }. 

In the case of planar scanning, a sampling step of half wavelength is enough to recover the visible portion of the far field spectrum and the radiation pattern of the AUT. Accordingly, a step of $\lambda/2$ represents the actual standard in planar near field antenna measurements.

Despite its simplicity, the half wavelength sampling does not take into account for the geometrical parameters of the configuration (the size and the shape of the source, the size of the measurement plane, and the distance between the source and the measurement plane); for such reason, it requires to acquire a redundant number of near field samples. Naturally, this negatively affects the acquisition time of the field samples which, especially at high frequencies, may become long. 

In this framework, two relevant points are: 
\begin{enumerate}
	\item to establish the minimum number of measurements that allows to acquire the near field without loss of information,
	\item to develop a sampling strategy of the near field that employs a non-redundant number of measurements. 
\end{enumerate}

As concerns the first point, the minimum number of measurements is equal to the \textit{number of degrees of freedom} of the near field. By definition, such a number represents the \textquotedblleft\textit{essential dimension}'' of the field space \cite{Barakat} \cite{Miller}. It can be evaluated by determining the number of relevant singular values of the radiation operator $T$ relating the source current $J$ with the near field $E$ \cite{NDF1} \cite{NDF2}.  

In reference to the second point, over the years different sampling strategies employing a reduced number of near field measurements have been proposed \cite{local1,local2,adaptive1,adaptive2,adaptive3,selection1,selection2,selection3,selection4,selection5,sparse1,sparse2,sparse3,other1,svd2,svd3,MDPI2020}. Among these, the sampling scheme described in \cite{MDPI2020} for a finite planar source is of particular interest since it exploits a number of near field measurements equal to the NDF. Such strategy estimates the sampling points of the near field $E$ by discretizing the continuous model $TJ=E$ in such a way that the discretized model shows the same singular values of the continuous one. More in detail, the problem of discretizing the radiation operator $T$ is cast as the discretization of the composed operator $TT^\dag$ where $T^\dagger$ stands for the adjoint operator. To this end, at first, a warping vector transformation mapping the two starting observation variables into two new ones is introduced to rewrite the kernel of $TT^\dag$ as a convolution and bandlimited function. After, the continuous operator $TT^\dag$ is discretized by exploiting the Shannon sampling theorem \cite{Khare} and a sampling series of the near field is derived. 

The vector warping transformation derived in \cite{MDPI2020} works well until the observation domain is not larger than the source domain. When this condition is breached, the derived warping transformation does not provide a good representation of the field outside the source domain and the resulting sampling series returns an under-sampled version of the field.
Accordingly, a more complicated warping transformation where each new variables depend on both starting variables should be introduced. To find such a transformation is a difficult task and it leads to a complicated no-factorized sampling point distribution. 

To overcome such a problem, a sub-optimal sampling strategy is proposed in \cite{Access2021}. Basically, the arrangement of points provided by the warping sampling scheme is thickened by employing an oversampling factor. In particular, since the oversampling is strictly required only outside the source domain, a spatially varying oversampling is introduced. The resulting sampling scheme employs a number of field samples slightly higher than the NDF but significantly lower than half-wavelength sampling. Hence, it is almost optimal in terms of number of measurements. Despite this, the arrangement of the sampling points over the measurement plane is non-uniform; this represents an issue since in many cases the control software of the near field probe is not conceived to move it with a non uniform step. 
Moreover, another issue of the sampling scheme in \cite{Access2021} is represented by the choice of the constants appearing in the oversampling factor since they depend on the geometrical parameters of the configuration in an unknown manner.

With the aim of avoiding the drawbacks of non-uniform sampling strategies, a uniform sampling scheme will be presented. The latter will be obtained with a proper choice of the oversampling functions whose structure is uniquely determined once the geometrical parameters of the configuration have been specified.

The paper is structured as follows. In section II, the radiation operator linking the planar current to the near field over a plane is introduced. In section III, the near field sampling scheme based on the warping technique is recalled from the literature. In section \ref{4a}, the warping sampling scheme is extended to the case of a measurement plane larger than the source domain. In section \ref{4b}, starting from the sampling criterion of section \ref{4a}, a novel uniform sampling criterion is derived. Finally, a section of numerical results based on experimental data is included to assess the performance of the new sampling scheme.  

\section{Geometry and mathematical preliminaries}
Consider the $3D$ geometry depicted in Figure 1 where
a planar antenna supported over a finite plane $SD=[-a,a]\times[-b,b]$ located at $z=0$ is shown.
\begin{figure} [htpb]
	\centering
	\includegraphics[scale=0.6]{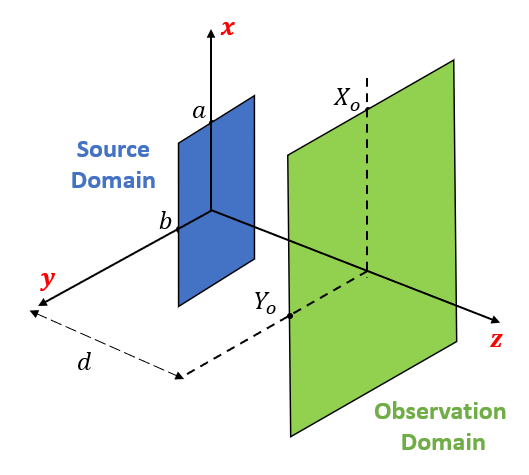}
	\caption{Geometry of the problem.}
\end{figure}

According to \cite{{NFFFT2}}, the electromagnetic field radiated by a planar antenna is equivalent to the one generated by a magnetic density current $\mathbf{J}(x,y) = J_x(x,y)\, \mathbf{\hat i_x}+J_y(x,y)\, \mathbf{\hat i_y}$.

Such density current radiates an electric field $\mathbf{E}$ whose tangential components $(E_x , E_y)$ are collected  over a finite plane $OD=[-X_o,X_o]\times[-Y_o,Y_o]$ located at $z=d$.

For the considered configuration, the components of the density current are related to the tangential components of the electric field by the scalar equations
\begin{equation} 
	E_y(\textbf{r})=T\, J_x(\textbf{r}') 
	\label{1}
\end{equation}
\begin{equation} 
	E_x(\textbf{r})=T\, J_y(\textbf{r}')
	\label{1bis}
\end{equation}
where 
\begin{itemize}
	\item $\textbf r'=(x',y',0)$ is a point of the source domain $SD$,
	\item $\textbf r=(x,y,d)$ is a point of the observation domain $OD$,
	\item $T: J\in L^2 (SD)\longrightarrow E\in L^2 (OD)$ is the radiation operator with $L^2(\cdot)$ denoting the class of square integrable functions over the set $(\cdot)$.
\end{itemize}
Since equations \eqref{1} and \eqref{1bis} are formally identical, in the following the simplified notation $E=TJ$ will be used to refer to both the equations. 

The radiation operator $T$ can be explicitly written as 
\begin{equation} 
	T\, J(\textbf{r}')=\int_{SD} h(\textbf{r},  \textbf{r}')\, e^{-j k |\textbf r-\textbf r'|} J(\textbf{r}')\, d\textbf r'
	\label{1tris}
\end{equation}
where $k$ denotes the wavenumber, and $h( \textbf{r},  \textbf{r}')$ is an amplitude term. Assuming $d$ large enough to neglect the evanescent waves contribution, the function $h( \textbf{r},  \textbf{r}')$ can be approximated as 
\begin{equation} 
	h( \textbf{r},  \textbf{r}')\approx- \dfrac{j k d}{4\pi |\textbf{r}-\textbf{r}'|^2  }.
\end{equation} 

As concerns the adjoint operator $T^\dagger : E\in L^2 (OD)\longrightarrow J\in L^2 (SD)$, it is given by

\begin{equation} 
	T^\dag\, E(\bold r)=\int_{OD} h^*(\textbf r, \textbf r')\, e^{j k |\textbf r-\textbf r'|} d\textbf r
\end{equation}
with $h^*(\textbf r, \textbf r')$ denoting the conjugate of $h(\textbf r, \textbf r')$.

Since the near field sampling proposed here is based on  a Shannon sampling representation of the left singular functions of the radiation operator, the singular system of $T$ is introduced. The latter is made of the triple $\{u_n, v_n, \sigma_n\}$ where 
\begin{itemize}
	\item $\{\sigma_n\}$ are non-decreasing scalars called \textit{singular values},
	\item $\{u_n\}$ and $\{v_n\}$ are two sets of orthonormal functions named respectively \textit{right} and \textit{left singular functions}. 
\end{itemize}
By definition, the singular system of $T$ satisfies the equations $Tu_n=\sigma_n v_n$ and   $T^\dagger v_n=\sigma_n u_n$ for each $n\in\mathbb{N}$. Accordingly, the $n-$th right singular function fulfills the eigenvalue problem $T^\dagger Tu_n=\sigma_n^2 u_n$ while the $n-$th left singular function is a solution of $TT^\dagger v_n=\sigma_n^2 v_n$.

\section{Near field sampling when $OD\subseteq SD$}
In this section, the near field warping sampling scheme working until $OD\subseteq SD$ ($X_o\leq a$ and $Y_o\leq b$) is recalled from \cite{MDPI2020}.

When the smallest size of the source is larger than the wavelength (\,$min\{a,b\}>>\lambda/2\pi$\,), the singular values of $T$ exhibit an abrupt decay at a critical index $N$ which identifies the \textit{number of degrees of freedom}. 
In such a case, the range of  $T$ is ``essentially" finite-dimensional; accordingly, the near field can be
represented by considering only the first $N$ left singular functions, i.e.
\begin{equation}
	E(\textbf r)\approx\sum_{n=1}^{N} c_n v_n(\textbf r)
	\label{6}
\end{equation} 

From Equation \eqref{6} it is evident that a field sampling criterion can be derived by a proper discretization of the left singular functions $\{v_1(\textbf r), ..., v_{N}(\textbf r)\}$. To find such discretization, the eigenvalue problem $TT^\dagger v_n=\sigma_n^2 v_n$ is considered. The latter can be expressed as
\begin{equation} 
	\int_{OD} H(\textbf{r},  \textbf{r}_o)\, v_n(\textbf r_o)\, d\textbf r_o=\sigma_n^2\, v_n(\textbf r)
	\label{8}
\end{equation}
where $\textbf r_o=(x_o,y_o,d)$ and 
\begin{equation}
	\hspace{-3pt}H(\textbf{r},  \textbf{r}_o)=\int_{SD} h^*(\textbf r_o, \bold r')\, h(\textbf{r},  \textbf{r}')\,   e^{j k \big(|\textbf r_o-\textbf r'|-|\textbf r-\textbf r'|\big)}  \, d\textbf r'.
\end{equation}
As shown in \cite{MDPI2020}, Equation \eqref{8}
can be approximated as
$$\hspace{9pt}\int_{\zeta(-Y_o)}^{\zeta(Y_o)}\int_{\eta(-X_o)}^{\eta(X_o)}  e^{-j\, k a\, [\gamma(\eta)-\gamma(\eta_o)]} e^{-j\,k b\, [\xi(\zeta)-\xi(\zeta_o)]} \ \ \ \ \ \ \ $$
\begin{equation}
	\begin{split}
		\hspace{-5pt}	sinc\,(ka\,(\eta-\eta_o))\, sinc\,(kb\,(\zeta-\zeta_o)) v_n(\eta_o,\zeta_o)\, d\eta_o\, d\zeta_o 
	\end{split}
	\label{12}
\end{equation}
$$\hspace{150pt} =\, \sigma_n^2 v_n(\eta,\zeta)$$
where the new variables $\eta=\eta(x)$, $\eta_o=\eta(x_o)$, $\zeta=\zeta(y)$, $\zeta_o=\zeta(y_o)$ are defined as
\begin{equation}
	\eta(x)=\dfrac{\sqrt{(x+a)^2+d^2}-\sqrt{(x-a)^2+d^2}}{2a}
\end{equation}
\begin{equation}
	\zeta(y)=\dfrac{\sqrt{(y+b)^2+d^2}-\sqrt{(y-b)^2+d^2}}{2b}
\end{equation}
and
\begin{equation}
	\gamma(x)=\frac{\sqrt{(x+a)^2+d^2}+\sqrt{(x-a)^2+d^2}}{2a}
\end{equation}
\begin{equation}
	\xi(y)=\frac{\sqrt{(y+b)^2+d^2}+\sqrt{(y-b)^2+d^2}}{2b}
\end{equation}
At this juncture, if the assignment
\begin{equation}
	\hat v_n(\eta,\zeta)= v_n(\eta,\zeta)\, e^{jk a\, \gamma(\eta)} e^{jk b\, \xi(\zeta)}
	\label{v_n_cappello1}
\end{equation}
is performed, the eigenvalue problem \eqref{12} can be rewritten as
\begin{equation}
	\begin{split}
		&\int_{\zeta(-Y_o)}^{\zeta(Y_o)}\int_{\eta(-X_o)}^{\eta(X_o)} sinc{((ka\, (\eta-\eta_o))}\, sinc{(kb\,(\zeta-\zeta_o))} \\& \hspace{90pt} \hat v_n(\eta_o,\zeta_o)\, d\eta_o d\zeta_o=\sigma_n^2\, \hat v_n(\eta,\zeta)
	\end{split}
	\label{14}
\end{equation}

Differently by \eqref{8}, the previous integral equation involves a convolution operator with a bandlimited kernel. Accordingly, also the eigenfunctions $\{\hat v_n\}$ are bandlimited with respect to $(\eta,\zeta)$ with a bandwidth equal to $k a$ and $k b$, respectively. For such reason, they can be discretized without loss of information with the step-lengths $\Delta \eta ={\pi}/(k a)$ and  $\Delta \zeta =\pi/(k b)$.
Moreover, the first $N$ eigenfunctions can be represented through the following truncated Shannon sampling expansion
\begin{equation}
	\hat{v}_n(\eta,\zeta)\approx \hspace{280pt} \vspace{-7pt}
	\label{v_n_cappello2}
\end{equation}
$$ \sum_{p=-N_1}^{N_1}\sum_{q=-N_2}^{N_2}\hat{v}_n(\eta_{p},\zeta_{q})\,sinc(k a\, \eta -p\pi)\,sinc(k b\, \zeta -q\pi)$$
where $\eta_{p}=p\, \pi/(ka)$, $\zeta_{q}=q\, \pi/(kb)$,
$N_1=[\frac{k a}{\pi}\, \eta(X_o)]$, $N_2=[\frac{k b}{\pi}\,\eta(Y_o)]$ with $[\,\cdot\,]$ denoting the integer part.
The scalar $N_1$ and $N_2$ have been chosen in such a way to consider only the samples $\{\hat v_n(\eta_p,\zeta_q)\}$ supported over the set $[\eta(-X_o),\eta(X_o)]\times[\zeta(-Y_o),\zeta(Y_o)]$. This choice makes negligible the error due to the truncation of the indexes $p$ and $q$ in \eqref{v_n_cappello2} since all the samples $\{\hat v_n(\eta_p,\zeta_q)\}$ falling outside the measurement plane are not relevant for the representation of $\hat v_n(\eta,\zeta)$ over the set $[\eta(-X_o),\eta(X_o)]\times[\zeta(-Y_o),\zeta(Y_o)]$.  

Now, if equation \eqref{v_n_cappello2} is substituted in \eqref{v_n_cappello1}, the following sampling representation comes out
\begin{equation}
	{v}_n(\eta,\zeta)=e^{-jk a\, \gamma(\eta)} e^{-jk b\, \xi(\zeta)}\sum_{p=-N_1}^{N_1}\sum_{q=-N_2}^{N_2}{v}_n(\eta_p,\zeta_q) \hspace{20pt}\vspace{-4pt}
	\label{16}
\end{equation}
$$\hspace{-5pt}e^{\,jk a\, \gamma(\eta_p)} e^{\,jk b\, \xi(\zeta_q)}\,sinc(k a\, \eta -p\pi)\,sinc(k b\, \zeta -q\pi)$$

By virtue of \eqref{6}, the near field $E$ admits a sampling representation perfectly equal to \eqref{16}. Moreover, since $\eta=\eta(x)$ and $\zeta=\zeta(y)$, it can be expressed in terms of the variables $(x,y)$ as below 
\begin{equation}
	\begin{split}
		&E\,(x,y)=e^{-jk a\, \gamma(x)} e^{-jk b\, \xi(y)}\sum_{p=-N_1}^{N_1}\sum_{q=-N_2}^{N_2}E\,(x_p,y_q) \ \ \ \ \ \ \\&
		e^{jk a\, \gamma(x_p)} e^{jk b\, \xi(y_q)}\,sinc\big(k a\, \eta(x) -p\pi\big)\,sinc\big(k b\, \zeta(y) -q\pi\big)\vspace{2pt}
	\end{split}
	\label{17}
\end{equation}
where $x_p$ and $y_q$ are solution of $\eta(x_p)=\eta_p$ and $\zeta(y_q)=\zeta_q$, respectively. Accordingly, it results that
\begin{equation}
	x_p=\eta_p\sqrt{a^2-\dfrac{d^2}{1-\eta_p^2}} \ \ \ \ \ \ \	y_q=\zeta_q\sqrt{b^2-\dfrac{d^2}{1-\zeta_p^2}}
	\label{punti}
\end{equation}

Equation \eqref{punti} returns the positions where the field samples must be collected.
Since the non-linear dependence of $\eta(x)$ and  $\zeta(y)$ respectively on $x$ and $y$, uniform sampling of $\eta(x)$ and  $\zeta(y)$ returns a non-uniform grid in $x$ and $y$ with a number of sampling points equal to
\begin{equation}
	N\approx \left(2\left[\dfrac{k a}{\pi} \eta(X_o)\right]+1\right)\left(2\left[\dfrac{k b}{\pi} \zeta(Y_o)\right]+1\right)
\end{equation}
Since $N$ is very close to the \textit{number of degrees of freedom} of the near field \cite{NDF}, the interpolation formula \eqref{17} exploits a non-redundant number of field measurements. In this sense, the described near field sampling criterion is optimal.

\section{Near field sampling when $OD\supseteq SD$}

The sampling scheme shown in Section III works until the observation domain ($OD$) is lower or equal than the source domain ($SD$). Hence, it can be exploited only for testing antennas with a broadside radiation pattern. 

When the main lobe of the field pattern is scanned away from boresight, in order to control the truncation error, the near field shall be collected over a measurement plane larger than the source domain. For such reason, this section aims at showing two measurement schemes of the near field suitable when $OD\supseteq SD$ (i.e., when $X_o > a$ and\,/\,or $Y_o > b$). In particular, in section \ref{4a} 
a recent non uniform sampling scheme based on warping technique is recalled by the literature. In section \ref{4b}, a novel uniform sampling criterion that overcomes all the drawbacks of non-uniform strategies is proposed.              

\subsection{Non-uniform field sampling} 
\label{4a}

When $OD\supseteq SD$, the sampling expansion \eqref{16} does not approximate well the field on the subset of $OD$  exceeding $SD$. In particular, according to local band considerations addressed in \cite{Access2021}, Equation \eqref{16} 
returns an under-sampled representation outside $SD$.
In this case, the approach developed in \cite{MDPI2020}
can be still applied but a couple of warping variables where each one depends on both $x$ and $y$ variables shall be used. This implies that the sampling along $x$ and $y$ is no more factorized since the arrangement of samples along $x$ will depend also on $y$ and vice versa.

Despite the non-factorized sampling is optimal in terms of number of samples, it introduces a complicated sampling scheme. This complication  can be overcome by introducing a uniform oversampling factor in the interpolation formula \eqref{16}. The latter reduces the representation error on the near field by using a slightly redundant number of samples (bigger than the number of degrees of freedom  but lower than the number of samples required by the half wavelength sampling). In particular, if the oversampling factors $\chi_1>1$ and $\chi_2>1$ are properly chosen, the left singular functions $\{v_1, ..., v_{N}\}$ are well approximated by
\begin{equation}
	v_n(\eta,\zeta)=e^{-jk a \gamma(\eta)} e^{-jk b \xi(\zeta)}\sum_{p=-M_1}^{M_1}\sum_{q=-M_2}^{M_2}v_n(\eta_p,\zeta_q) \ \ \ \ \ \ \ \  \vspace{-1pt}
	\label{v_n_oversampling1}
\end{equation}	
$$\hspace{-10pt}e^{jk a\, \gamma(\eta_p)} e^{jk b\, \xi(\zeta_q)}\,sinc\big(\chi_1 k a\, \eta -p\pi\big)\,sinc\big(\chi_2 k b\, \zeta -q\pi\big)$$
where $\eta_p=p\pi/(\chi_1 k a)$, $\zeta_q=q\pi/(\chi_2 k b)$, $M_1=[\frac{\chi_1 k a}{\pi}\, \eta(X_o)]$ and $M_2=[\frac{\chi_2 k b}{\pi}\, \zeta(Y_o)]$.

It is worth nothing that, for a correct representation of the left singular functions, the oversampling is not mandatory in the region where $|\eta|<\eta(a)$ and $|\zeta|<\zeta(b)$ since in this region the considerations of Section III can be applied. On the contrary, it is strictly necessary for $|\eta|>\eta(a)$ and $|\zeta|>\zeta(b)$.

Based on these considerations, in \cite{Access2021} a \textit{spatially varying oversampling} is proposed with the aim to re-arrange the sampling points by  
thickening the sampling step only outside $SD$. The spatially varying oversampling is realized by choosing two oversampling factors $\chi_1(x)$ and $\chi_2(y)$  as below
\begin{equation}
	\chi_1(x)=1-(1-\alpha_1)sin^p\left(\dfrac{\pi}{2X_o}x\right)
	\label{chi_1}
\end{equation}

\begin{equation}
	\chi_2(y)=1-(1-\alpha_2)sin^p\left(\dfrac{\pi}{2Y_o}y\right)
	\label{chi_2}
\end{equation}
Such factors are proximal to 1 for $|x|\leq a$ and $|y|\leq b$, instead, they rise monotonically for $\,a<|x|\leq X_o\,$ and $b<|y|\leq Y_o$. In this way, the sampling rate is kept at the Nyquist limit in the region of the measurement plane located in front of the source domain while it is gradually increased at the edges of the measurement plane.

When a spatially varying oversampling is adopted, the first $N$ left singular functions of the radiation operator can be still expressed trough a truncated Shannon sampling series. Indeed, after the introduction of the variables
\begin{equation}
	\hat{\eta}(x)=\chi_1(x)\,\eta(x) \ \ \ \ \ \ \ \ \hat{\zeta}(y)=\chi_2(y)\,\zeta(y)
\end{equation}
the left singular functions $\{v_n\}_{n=1}^{N}$ can be written as
\begin{equation}
	\hspace{-2pt}	v_n\,(\hat\eta,\hat\zeta)=e^{-jk a\, \gamma(\hat \eta)} e^{-jk b\, \xi(\hat \zeta)}\sum_{p=-M_1}^{M_1}\sum_{q=-M_2}^{M_2}v_n\,(\hat\eta_p,\hat\zeta_q)\vspace{-8pt}
	\label{v_n_oversampling1}
\end{equation}
$$e^{jk a\, \gamma(\hat\eta_p)} e^{jk b\, \xi(\hat\zeta_q)}\,sinc\big(k a\, \hat\eta -p\pi\big)\,sinc\big(k b\, \hat\zeta -q\pi\big)$$
with $\hat\eta_{p}=p\, \pi/(ka)$, $\hat\zeta_{q}=q\, \pi/(kb)$,
$M_1=[\frac{k a}{\pi}\, \hat\eta(X_o)]$, and $M_2=[\frac{k b}{\pi}\,\hat\eta(Y_o)]$. \vspace{2pt}
As a consequence, the near field $E(x,y)$  is given by 
\begin{equation}
	\begin{split}
		E&\,(x,y)=e^{-jk a\, \gamma(x)} e^{-jk b\, \xi(y)}\sum_{p=-M_1}^{M_1}\sum_{q=-M_2}^{M_2}E\,(x_p,y_q) \\& \hspace{-12pt}
		e^{jk a\, \gamma(x_p)} e^{jk b\, \xi(y_q)}\,sinc\big(k a\, \hat\eta(x) -p\pi\big)\,sinc\big(k b\, \hat\zeta(y) -q\pi\big)
	\end{split}
	\label{E_n_oversampling1}
\end{equation}
where  $x_p$ and  $y_q$ are solution of 
\begin{equation}
	\hat\eta(x_p)=\chi_1(x_p)\,\eta(x_p)=p\dfrac{\pi}{ka}
	\label{trasf1}
\end{equation}
\begin{equation}
	\hat\zeta(y_q)=\chi_2(y_q)\,\zeta(y_q)=q\dfrac{\pi}{kb}.
	\label{trasf2}
\end{equation}
Accordingly, following this approach, the variables to be uniformly sampled are not $\eta$ and $\zeta$ but $\hat\eta$ and $\hat\zeta$.

\subsection{Drawbacks of non-uniform methods and introduction of a novel uniform sampling}

\label{4b}
\begin{figure}[t]
	\centering
	\includegraphics[scale=0.63]{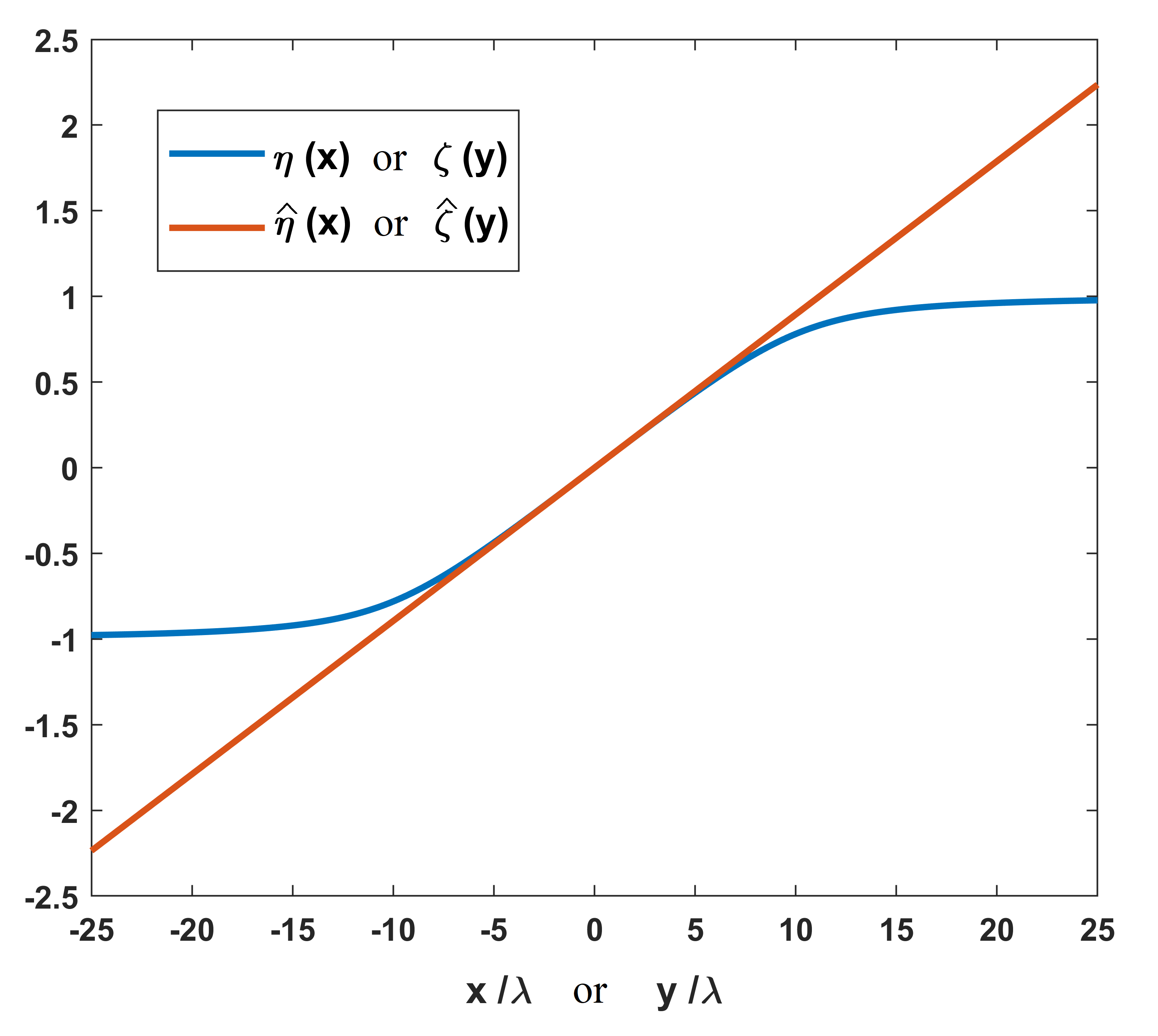}
	\caption{Diagrams of $\eta(x)$, $\hat\eta(x)$ and $\zeta(y)$, $\hat\zeta(y)$ when $a=b=10\lambda$  and $d=5\lambda$.}
	\label{fig2}
\end{figure}

Despite the introduction of the spatially varying oversampling factors described in section \ref{4a} is a sub-optimal strategy that allows achieving a better trade-off between the representation error and the number of sampling points than the one based on uniform oversampling, the proposed strategy still shows two drawbacks.

The first one concerns the non-uniform arrangement of the near field samples over the measurement plane which is due to the non-linear dependence of $\hat\eta(x)$ and  $\hat\zeta(y)$ on $x$ and $y$, respectively. Indeed, although the non-uniform sampling of the near field is possible in principle, the control software of many near field scanner systems does not give the possibility to move the field probe with a non-uniform step length. For such reason, non-uniform near field measurements are difficult to perform without updating the scanning system. 

The second drawback concerns the choice of the constants $\alpha_1$, $\alpha_2$ and $p$ appearing in the oversampling factors \eqref{chi_1} and \eqref{chi_2}. Such constants depend on the geometrical parameters of the configuration $a,b,X_o,Y_o,d$ in an unknown manner; accordingly, they must be properly chosen case by case.

With the aim to overcome these issues, a uniform sampling scheme of the near field is now proposed. As will be more clear later, such a strategy needs a number of measurements slightly higher than the non-uniform sampling scheme of section $IV A$ but significantly lower than the classical half wavelength sampling leading to a good trade-off between the acquisition time of the near field samples and the simplicity of the scanner system.

It is worth highlighting that the non-uniformity of the sampling schemes recalled above derives from the non-linear behavior of the warped variables in terms of the spatial variables $x$ and $y$. Indeed, due to such non-linearity, the uniform sampling in the warped variables is mapped into a non-uniform sampling in the original spatial variables. 

Moved by this reasoning and inspired by the spatially varying oversampling approach, our idea is to introduce two spatially varying oversampling factors that compensate for the non-linear behavior of the original warped variables $\eta(x)$ and $\zeta(y)$ in such a way that the resulting sampling in the original spatial variables $x$ and $y$ will be uniform. 
In particular, in the uniform sampling scheme proposed here, the oversampling factors $\chi_1(x)$ and $\chi_2(y)$ will be chosen in such a way that 
\begin{enumerate}
	\item  the new warped variables 
	$\hat \eta(x)=\chi_1(x)\, \eta(x)$ and $\hat \zeta(y)=\chi_2(y) \zeta(y)$ are linear functions; 
	
	\item the oversampling functions $\chi_1(x)$ and $\chi_2(y)$ are greater or equal than $1$ over all the observation domain $OD$. 
\end{enumerate}

The fulfillment of the first condition ensures that a uniform sampling of the warped variables $\hat \eta(x)$ and $\hat \zeta(y)$ is mapped into a uniform sampling in the original spatial variables $x$ and $y$. Instead, the fulfillment of the second condition allows to keep the representation error of the near field coming from the proposed uniform sampling comparable with the representation deriving from the half-wavelength sampling and non-uniform sampling of section IV A.

\begin{figure}[t]
	\includegraphics[scale=0.10]{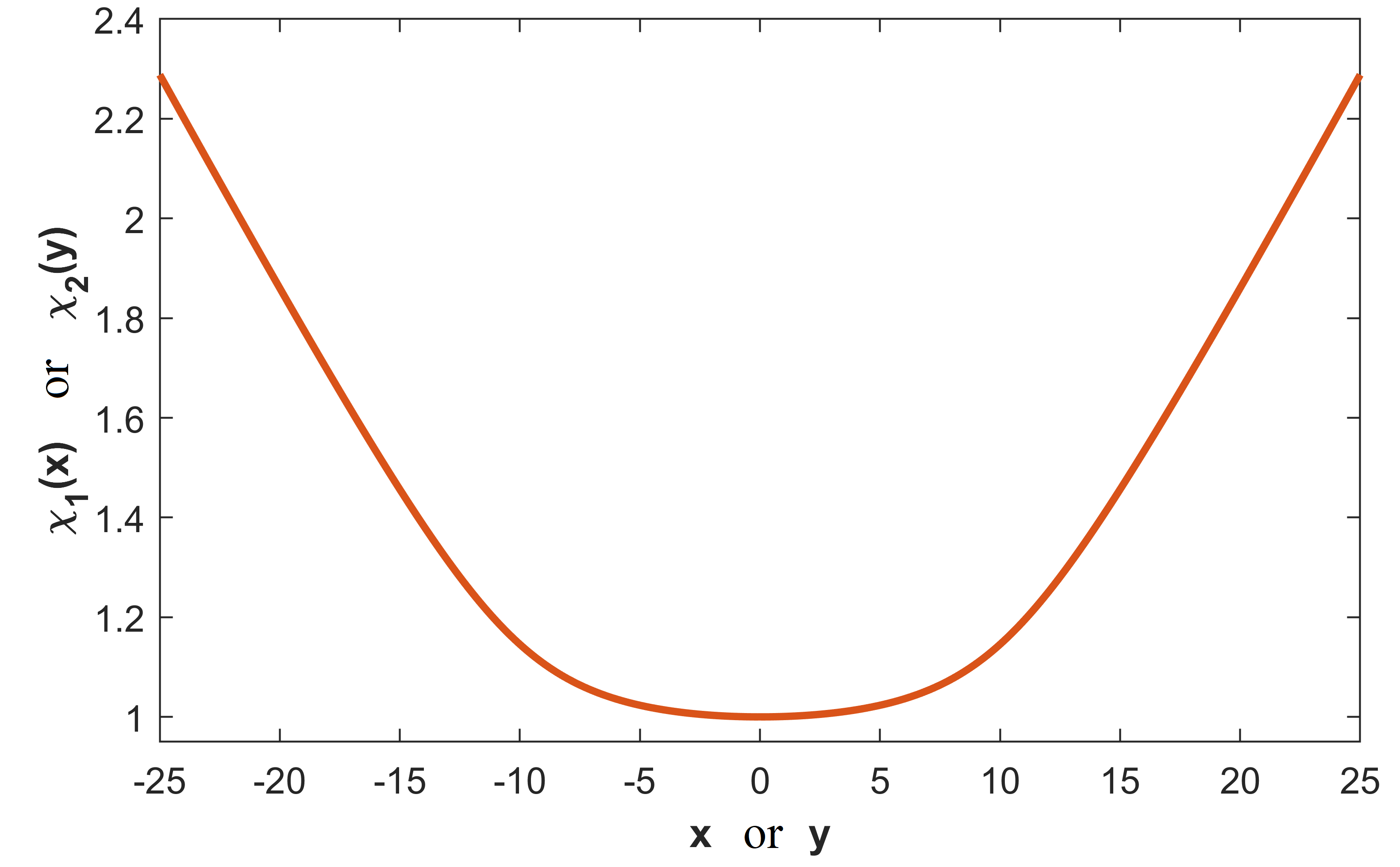}
	\caption{Diagrams of $\chi_1(x)$ and $\chi_2(y)$ when $a=b=10\lambda$ and $d=5\lambda$.}
	\label{fig3}
\end{figure}

To obtain the desired linear expression of the warped variables  $\hat \eta(x)$ and $\hat \zeta(y)$, the Taylor expansion at the first order of $ \eta(x)$ and  $ \zeta(y)$ around the points $x=0$ and $y=0$  is considered.  From such expansions, it comes out that  
\begin{equation}
	\hat \eta\,(x)=\left.\dfrac{d\eta}{dx}\right|_{x\,=\,0\,} x=\dfrac{1}{\sqrt{a^2+
			d^2}}\,x 
	\label{trasf11}
\end{equation}
\begin{equation}
	\hat \zeta(y)=\left.\dfrac{d\zeta}{dy}\right|_{y\,=\,0\,} y=\dfrac{1}{\sqrt{b^2+d^2}}\,y
	\label{trasf22}
\end{equation}
Since the spatially varying oversampling factors $\chi_1(x)$ and $\chi_2(y)$ shall be in agreement with \eqref{trasf11} and \eqref{trasf2}, it results that
\begin{equation}
	\hspace{-5pt}\chi_1(x)=\dfrac{x}{\sqrt{a^2+d^2
	}}\dfrac{2a}{ \sqrt{(x+a)^2+d^2}-\sqrt{(x-a)^2+d^2}}
	\label{ov11}
\end{equation}
\begin{equation}
	\hspace{-5pt}	\chi_2(y)=\dfrac{y}{\sqrt{b^2+d^2
	}}\dfrac{2b}{ \sqrt{(y+b)^2+d^2}-\sqrt{(y-b)^2+d^2}}
	\label{ov22}
\end{equation}

In Figure \ref{fig2} the diagrams of the warping variables $\eta(x)$, $\hat\eta(x)$ and $\zeta(y)$, $\hat\zeta(y)$ are sketched. As can be seen from the figure, it results that $|\hat \eta(x)|\geq |\eta(x)| \ \forall x\in [-X_o,X_o]$ and $|\hat \zeta(y)|\geq |\eta(y)| \ \forall y\in [-Y_o,Y_o]$. This implies that the choice of $\chi_1(x)$ and $\chi_2(y)$ ensuring a linear behavior of $\hat \eta(x)$ and $\hat \zeta(y)$ also guarantees $\chi_1(x)\geq 1 \ \forall x\in [-X_o,X_o]$ and $\chi_2(y)\geq 1 \ \forall y\in [-Y_o,Y_o]$. Hence, the uniform sampling proposed in this section is really an over-sampled version of the sampling criterion described in Section III.

The plot of $\chi_1(x)$ and $\chi_2(y)$ is shown in Figure \ref{fig3}. From such plots, it is evident that $\chi_1(x)\approx 1$ for  $|x|\leq a$ while $\chi_1(x)> 1$ for  $|x|> a$. Similarly, $\chi_2(y)\approx 1$ for  $|y|\leq b$ while $\chi_2(y)> 1$ for  $|y|> b$. This allows us to state that also the uniform sampling scheme employs a number of field measurements proximal to the minimum in the portion of the measurement plane located in front of the source. Instead, at the edges of the observation domain, the sampling rate is not decreased but maintained constant.

\begin{figure}[t]
	\centering 
	\includegraphics[scale=0.11]{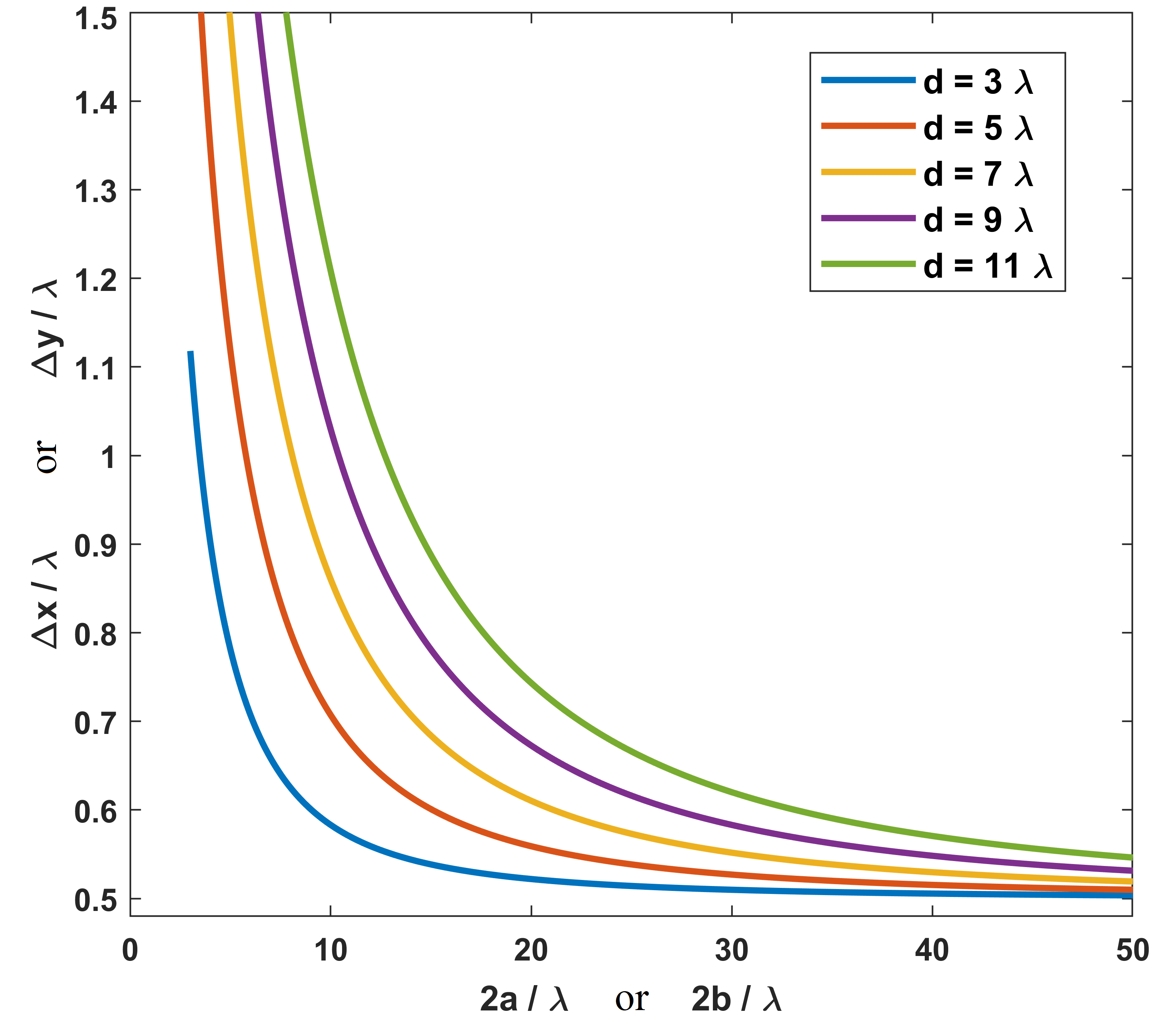}
	\caption{Diagram of the sampling step along $x$ (or $y$) in terms of the source length along $x$ (or $y$) for different value of the distance $d$.}
	\label{fig4}
\end{figure}

At this juncture, if equations \eqref{trasf11} and \eqref{trasf22} are considered, the general near field interpolation formula \eqref{E_n_oversampling1} can be rewritten as below
\begin{equation}
	\begin{split}
		\hspace{-2pt}E&(x,y)=e^{-jk a \gamma(x)} e^{-jk b \xi(y)}\sum_{p=-M_1}^{M_1}\sum_{q=-M_2}^{M_2}E(x_p,y_q)\, e^{jk a \gamma(x_p)} \\& \hspace{-5pt}
		e^{jk b\, \xi(y_q)}\,sinc\left(\dfrac{k a}{\sqrt{a^2+d^2}}\, x -p\pi\right)\,sinc\left(\dfrac{kb}{\sqrt{b^2+d^2}}\,y -q\pi\right)
	\end{split}
	\label{E_n_oversampling2}
\end{equation}
where 
\begin{itemize}
	\item $M_1=[\frac{k a}{\pi}\, \hat\eta(X_o)]=\left[2 aX_o/(\lambda\sqrt{a^2+d^2})\right]$,\vspace{1pt} 
	\item $M_2=[\frac{k b}{\pi}\,\hat\eta(Y_o)]=\left[2bY_o/(\lambda\sqrt{b^2+d^2})\right]$.
\end{itemize}

The exact coordinates $\{x_p\}$ and $\{y_q\}$ of the sampling points can be easily derived by substituting the expression of $\hat\eta(x)$ and $\hat\zeta(y)$ provided by \eqref{trasf11} and \eqref{trasf22} in \eqref{trasf1} and \eqref{trasf2}, respectively.  By doing this and expressing the wavenumber in terms of the wavelength ($k=2\pi/\lambda$), it comes out that
\begin{equation}
	x_p=p\dfrac{\lambda}{2}\dfrac{\sqrt{a^2+d^2}}{a} \ \ \ \ \ \ \ \ \ y_q=q\dfrac{\lambda}{2}\dfrac{\sqrt{b^2+d^2}}{b}.
	\label{trasf111}
\end{equation}
$\forall p\in\{-M_1,\ldots,M_1\}$ and $\forall q\in\{-M_2,\ldots, M_2\}$.
Hence, the sampling steps $\Delta x$ and $\Delta y$ are given by
\begin{equation}
	\Delta x=\dfrac{\lambda}{2}\dfrac{\sqrt{a^2+d^2}}{a} \ \ \ \ \ \ \ \ \ \Delta y=\dfrac{\lambda}{2}\dfrac{\sqrt{b^2+d^2}}{b}.
	\label{steps}
\end{equation} 
From the previous equation, it is evident that the distance between the field samples is always greater than $\lambda/2$; accordingly, the required number of field measurements 
\begin{equation}
	\hspace{-5pt}	M=\left(2\left[\dfrac{2X_o}{\lambda}\dfrac{a}{\sqrt{a^2+d^2}}\right]+1\right)\left(2\left[\dfrac{2Y_o}{\lambda}\dfrac{b}{\sqrt{b^2+d^2}}\right]+1\right)
	\label{M}
\end{equation}
is always lower than the half-wavelength sampling. In particular, the steplengths $\Delta x$ or $\Delta y$ of the proposed sampling scheme are significantly larger than $\lambda/2$ 
especially when the semi-extension of source domain edges ($a$ or $b$) is little or comparable with respect to the distance $d$. This is also confirmed by Figure \ref{fig4} which illustrates the sampling step $\Delta x$ (or $\Delta y$) in terms of the source extension along $x$ (or $y$).

Equations \eqref{E_n_oversampling2} and \eqref{trasf111}  represent the basis of the proposed sampling scheme and the main results of this article. The main peculiarity of the proposed strategy for near field measurements is that it employs a number of field samples $M$ a few larger than the non-uniform sampling of Section \ref{4a} with the simplicity and the flexibility of the uniform half-wavelength sampling. Accordingly, it merges the main advantages of the two sampling schemes by avoiding their drawbacks. 

Finally, it is worth computing also the percentual reduction $R$ of field samples with respect to half-wavelength sampling. The latter is given by
\begin{equation}
	R=\left(1-\dfrac{M}{M_{\frac{\lambda}{2}}}\right) 100\%=\left(1-\dfrac{ab}{\sqrt{a^2+d^2}\sqrt{b^2+d^2}}\right) 100\%
\end{equation} 
where $ M_{\frac{\lambda}{2}} = (2\, [2X_o/\lambda] +1) (2 \,[2Y_o/\lambda] +1)$ is the number of field samples required by the half-wavelength sampling.

\section{Uniform sampling validation by\\ experimental field data}

\begin{figure*}[!t]
	\centering
	\includegraphics[scale=0.092]{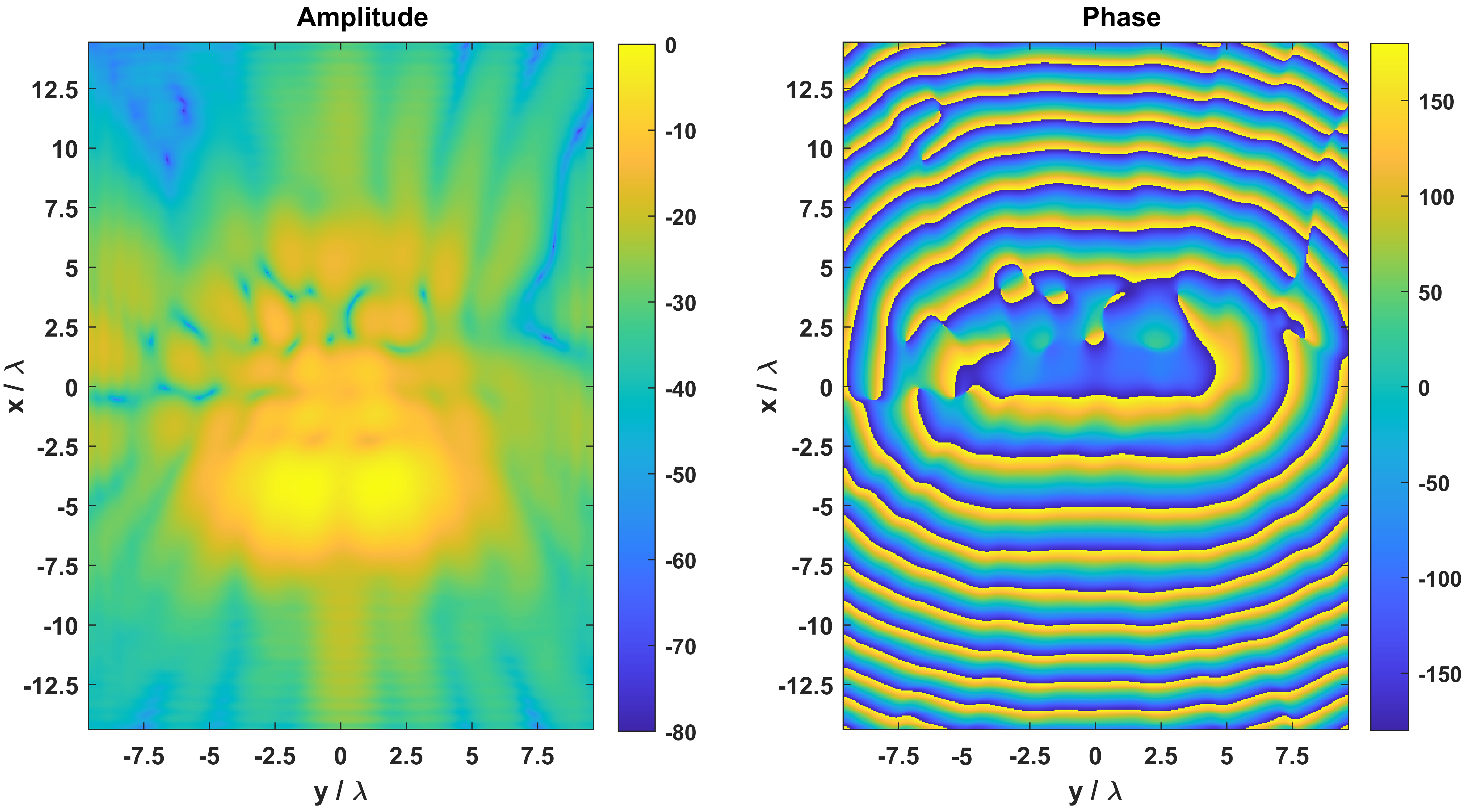}
	\caption{Near field obtained by the interpolation of measurements collected with $\Delta x=0.48\lambda$, $\Delta y= 0.48 \lambda$.}
	\label{fig5}
\end{figure*}

\begin{figure*}[!t]
	\centering
	\includegraphics[scale=0.092]{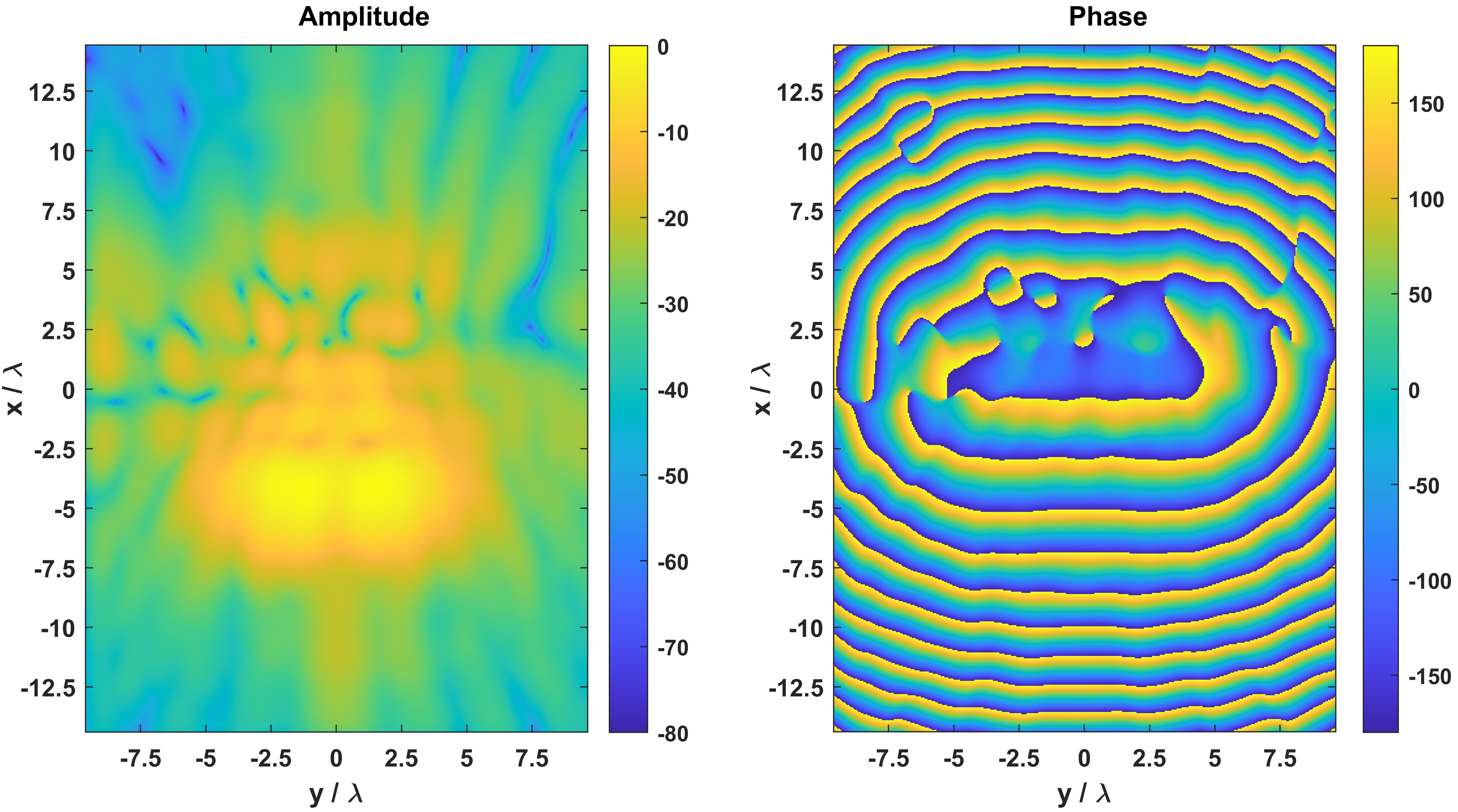}
	\caption{Near field obtained by the interpolation of measurements collected with $\Delta x=1.26\lambda$, $\Delta y = 0.86 \lambda$.}
	\label{fig6}
\end{figure*}

\begin{figure*}[!t]
	\centering
	\includegraphics[scale=0.092]{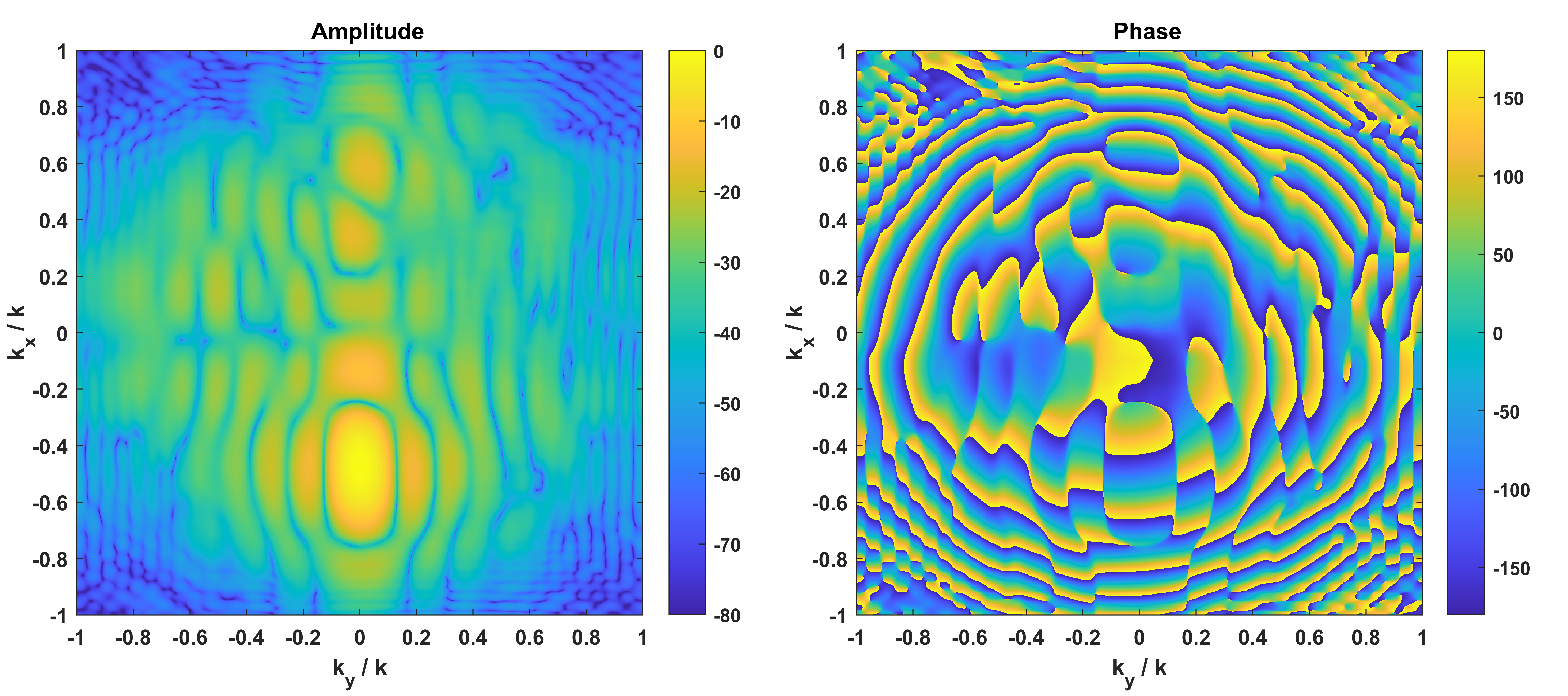}
	\caption{Far field spectrum obtained by the near field samples collected with the steplengths $\Delta x = 0.48\lambda$ and $\Delta y = 0.48\lambda$  }
	\label{fig7}
\end{figure*}

\begin{figure*}[!t]
	\centering
	\includegraphics[scale=0.57]{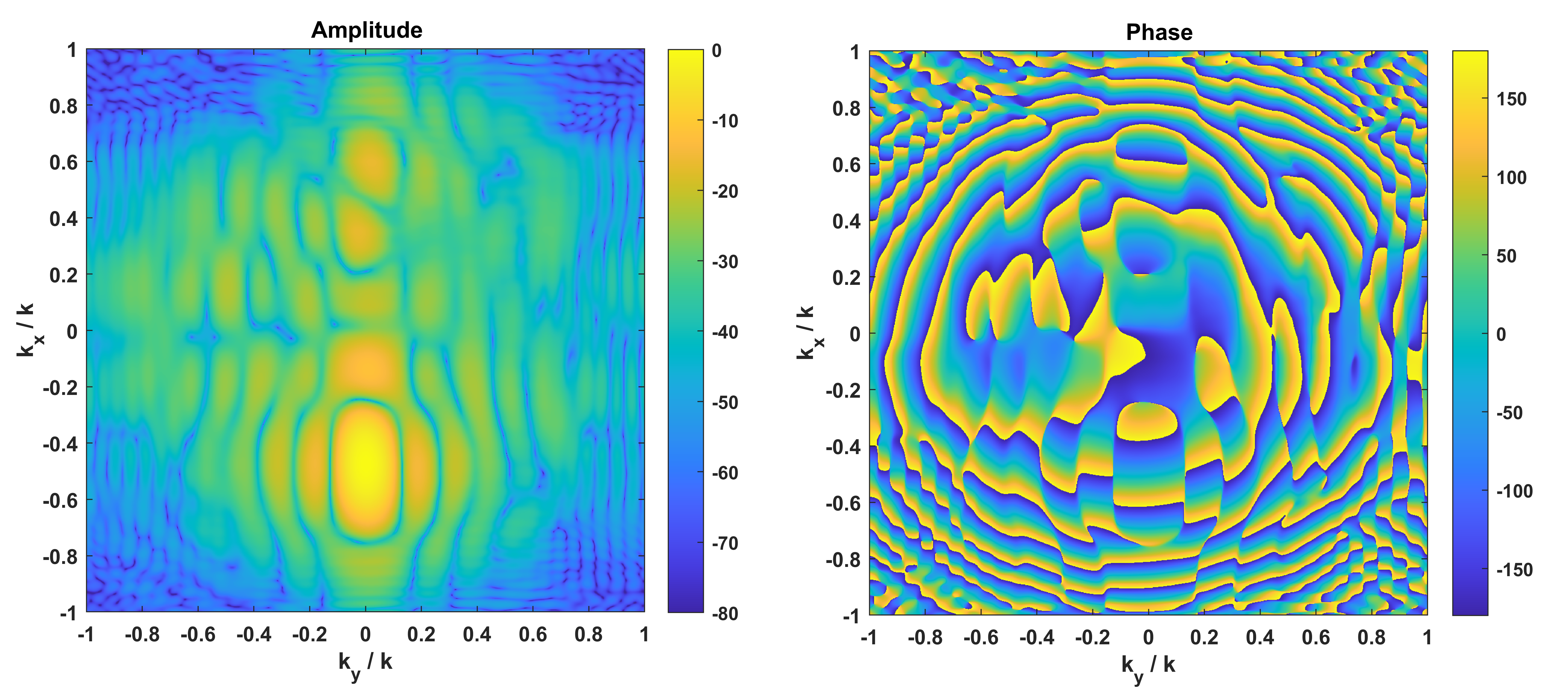}
	\caption{Far field spectrum obtained by the near field samples collected with the steplengths $\Delta x = 1.26\lambda$ and $\Delta y = 0.86\lambda$  }
	\label{fig8}
\end{figure*}

\begin{figure*}[htpb]
	\centering
	\includegraphics[scale=0.67]{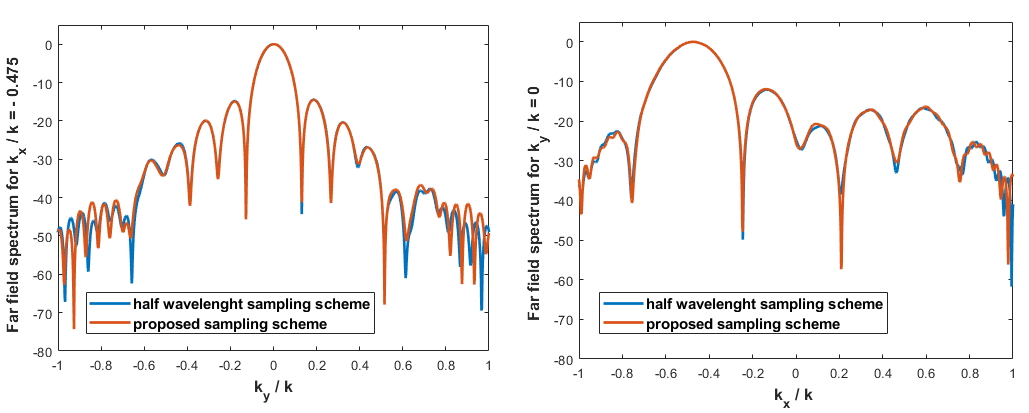}
	\caption{Comparison between the main cuts of the far field spectrum obtained the half wavelength sampling and the proposed sampling scheme.}
	\label{fig9}
\end{figure*}

In this section, the uniform sampling scheme proposed in section IV B is corroborated by exploiting some experimental field data provided by MBDA Italy. 

A rectangular array of Vivaldi antennas arranged over the source domain according to a triangular lattice
has been considered and tested.  
The array support is a rectangle whose edges have a semi-extension $a=3\lambda$, $b=5\lambda$.

The tangential component of the near field radiated by such array has been measured over a plane at a distance $d=7\lambda$. The semi-extension of the measurement plane along the $x$ and the $y$ axes are $X_o=14.5\lambda$ and $Y_o=9.75\lambda$, respectively.

In order to assess the performances of the proposed sampling scheme, the near field radiated by the array has been measured by employing both the classical step length of half-wavelength and the sampling steps provided by \eqref{steps}. Then, the near field retrieved by collecting data according to the standard half-wavelength sampling is compared with that obtained by exploiting the proposed uniform measurement strategy. Finally, also a comparison between the far field reconstructions is performed. 

In the first test case, the near field has been collected by employing the classical half-wavelength sampling ($\Delta x = 0.48 \lambda$, $\Delta y = 0.48 \lambda$). Accordingly, the number of field samples collected along the $x$-axis are $M_x=61$ while those collected along the $y$-axis are $M_y=41$. The total number of field samples is $M_{\lambda/2}=2501$.\\
To reconstruct the near field by its samples, the field measurements have been interpolated with the classical interpolation formula 

\begin{equation}
	\begin{split}
		&E(x,y)=\sum_{p=-30}^{30}\sum_{q=-20}^{20} E\left(p {\Delta x},q {\Delta y}\right)\,
		sinc\left(\dfrac{\pi}{\Delta x} x -p\pi\right)\\&\,sinc\left(\dfrac{\pi}{\Delta y}\,y -q\pi\right)
	\end{split}
	\label{classica}
\end{equation}
The amplitude in dB and the phase in degrees of the near field reconstructed by \eqref{classica} are sketched in Figure \ref{fig5}.

In the second test case, the near field has been measured by employing the novel uniform sampling introduced in section \ref{4b}. In such a case, the sampling steps deriving from \eqref{steps} are $\Delta x = 1.26\lambda$ and $\Delta y = 0.86 \lambda$ (note that the step length is smaller along the direction where the array under test is larger). Accordingly, the number of field samples collected along the $x$ and $y$ axes is respectively $N_x=24$ and $N_y=23$ while the total number of field measurements $M$ is  $552$.\\
The near field has been recovered by employing the interpolation formula \eqref{E_n_oversampling2}. The result of this interpolation is shown  
in Figure \ref{fig6} which sketches the amplitude in dB and the phase in degrees of the near field reconstruction. 

With the aim to evaluate the mismatching of the near field reconstructions shown in Fig. \ref{fig5} and \ref{fig6}, the relative root mean squared error ($rRMSE$) has been evaluated. In particular, it results that
\begin{equation}
	rRMSE=\dfrac{||E-E_{\lambda/2}||^2}{||E_{\lambda/2}||^2}=7.80\cdot10^{-4}
\end{equation}
where $||\cdot ||$ stands for the Euclidean norm. Accordingly, in the considered test cases, the proposed sampling scheme achieves the same reconstruction of the near field with a percentual reduction $R$ of field measurements equal to $77.93\%$.

Since in the framework of antenna testing the interest is in the far field of the antenna under test, the far field of the considered array has also been evaluated. 

In Figure \ref{fig7}, the far field spectrum obtained by near field samples collected with $\Delta x=0.48\lambda $ and $\Delta y = 0.48\lambda$ is shown. The figure has been obtained by computing directly the Fast Fourier Transform of the measured field samples. 

In Figure \ref{fig8}, the far field spectrum computed by the near field measurements collected with the step lengths  $\Delta x=1.26\lambda $ and $\Delta y = 0.86\lambda$ is sketched. In such a case, the far field spectrum has been evaluated in two steps. The first step is the resampling of the near field in Fig. \ref{fig6} with a step length of $\lambda/2$. Then, the Fast Fourier Transform of the synthetic field samples has been computed to obtain the far field spectrum. 

By comparing Figure \ref{fig7} and \ref{fig8}, it is evident that the reconstructions of far field spectrum deriving by the two sampling schemes are essentially the same. The accuracy in the reconstruction of the far field spectrum can also be noted by Figure \ref{fig9} which compares the main cuts of the far field spectrum for both the classical and the novel uniform sampling strategies.  

\section{Conclusions}

In this paper, a new sampling scheme of the near field radiated by planar source has been proposed and assessed. Such sampling scheme allows a reduction of the testing time for all the planar antennas whose edges have a length in the range $[3\lambda, 25\lambda]$. In particular, for all antennas of such dimensions, the proposed sampling scheme maintains the same performance of the half-wavelength sampling with a smaller number of near field measurements. 
Moreover, it employs a number of field samples slightly higher than non-uniform sampling schemes by avoiding the issues of non-uniform measurement strategies (not all the near field scanners are able to move the probe according to a non-uniform step).

Finally, it is worth remarking that for antennas greater than $25\lambda$, the proposed sampling scheme becomes similar to the half-wavelength sampling; instead, for antennas  smaller than $3\lambda$, the proposed uniform sampling may still be used if a slight oversampling factor is introduced. However, the quantification of such oversampling constant is out of the scope of this article and it will be demanded in a future study.

Other future developments concern the extension of the proposed uniform sampling to the case of amplitude-only measurements.

\begin{IEEEbiography}[{\includegraphics[width=1in,height=1.18in]{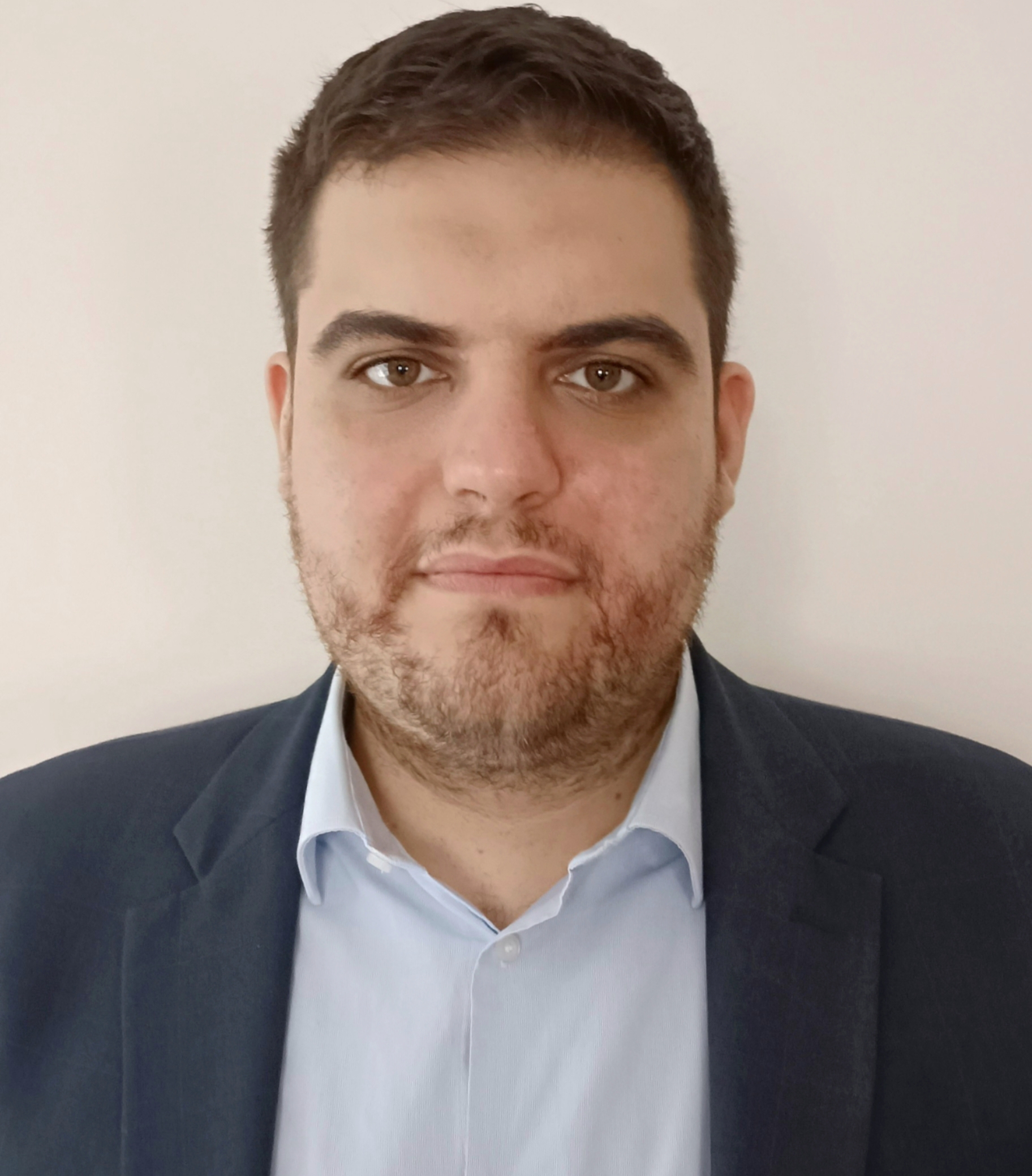}}]{Raffaele Moretta} (S' 2018 -- M' 2021) was born in Caserta (Italy) in March 1989.
	
He received the Laurea (summa cum laude) in Electronic Engineering and the Ph.D degree in Electronic and Computer Science engineering from the University of Campania ``Luigi Vanvitelli" in 2018 and 2021, respectively. 
			                    
At the end of 2021, he joined the Electromagnetic Fields Group of the University of Campania "Luigi Vanvitelli" as Postdoctoral Researcher.

After a short experience in Avio as avionic system engineer, he moved to MBDA Italy where he is now an antenna engineer.	
	                                                              His current research interests include inverse problems in electromagnetics with particular attention to phase retrieval,  near field measurement techniques, antenna diagnostics and synthesis.
	                                                              
Dr. Moretta is a member of the Institute of Electrical and Electronics Engineers (IEEE) and of the Italian Society of Electromagnetism (SIEM). Moreover, in February 2021 he was under consideration by the committee of the IEEE Antennas Propagation Society for the R.W.P. King Award 2020.
\end{IEEEbiography}

\begin{thebibliography}{00}
\bibitem{compact1} R.C. Johnson, H.A. Ecker, R.A. Moore, ``Compact range techniques and measurements,'' \textit{IEEE Trans. Antennas Propag.}, vol. 17, pp. 563--576, Sept. 1969.


\bibitem{near1} A. D. Yaghjian, ``An overview of near-field antenna measurements,'' \textit{IEEE Trans. Antennas Propag.}, vol. 34, no. 1, pp. 30-45, 1986.


\bibitem{NFFFT1} J. C. Bennett, E. P. Schoessow, ``Antenna near-field\,/\,far-field transformation using a plane-wave synthesis technique,''  \textit{Proc. IEE}, vol. 125, no. 3, pp. 179-184, March 1978.

\bibitem{NFFFT2} P. Petre, T. K. Sarkar,  ``Planar near-field to far field transformation using an equivalent magnetic current approach,'' \textit{IEEE Trans. Antennas Propag.}, vol. 40, pp. 1348-1356, 1992.

\bibitem{NFFFT3} A. D. Yaghjian, ``Upper bound errors in far-field antenna parameters determined from planar near-field measurements'', Nat. Bur. Stand., Boulder, CO, Tech. Rep. NBS Tech. Notes 667, 1975.

\bibitem{NFFFT4} Y. Rahmat-Samii, V. Galindo-Israel, R. Mittra, ``A plane-polar approach for far field construction from near-field measurements'' \textit{IEEE Trans. Antennas Propag.}, vol. 28, no. 2, pp. 216-230, March 1980. 

\bibitem{NFFFT5} O. M. Bucci, C. Gennarelli, C. Savarese, ``Fast and accurate near-field far-field transformation by sampling interpolation of plane polar measurements,'' \textit{IEEE Trans. Antennas Propag.}, vol. 39, pp. 48-55, Jan. 1991.

\bibitem{NFFFT6} G.F. Ricciardi, W.L. Stutzman, ``A near-field to far-field transformation for spheroidal geometry utilizing an eigenfunction expansion,'' \textit{IEEE Trans. Antennas Propag. }, vol. 52, no. 12, Dec. 2004

\bibitem{NFFFT7} M.A. Qureshi, C.H. Schmidt, T.F. Eibert, ``Efficient near-field far-field transformation for nonredundant sampling representation on arbitrary surfaces in near-field antenna measurements'' \textit{IEEE Trans. Antennas Propag.}, vol. 61, pp. 2025–2033, 2013.

\bibitem{NFFFT8} G. Giordanengo, M. Righero, F. Vipiana, G. Vecchi, M. Sabbadini, ``Fast antenna testing with reduced near field sampling,'' \textit{IEEE Trans. Antennas Propag.}, vol. 62, no. 5, pp. 2501--2513, 2014.

\bibitem{NFFFT9} F. D'Agostino, F. Ferrara, C. Gennarelli, R. Guerriero, ``Pattern reconstruction of 3-D modular antennas by means of a non-redundant near-field spherical scan,'' \textit{Electronics}, vol. 11, no. 13, art. no. 2060, 2022.

\bibitem{NFFFT10}
D. R. Prado, M. Arrebola, M. R. Pino, F. Las-Heras, "An Efficient Calculation of the Far Field Radiated by Non-Uniformly Sampled Planar Fields Complying Nyquist Theorem," \textit{IEEE Trans. Antennas Propag.}, vol. 63, no. 2, pp. 862-865, Feb. 2015.


\bibitem{compact_near_field} D.W. Hess, J.J. Tavormina, ``Verification testing of spherical near-field algorithm and comparison to compact range measurements,'' \textit{Int. Symp. Dig. IEEE Antennas Propag. Soc.}, vol. 1, pp. 242--271, Los Angeles, June 16--19, 1981.


\bibitem{planar1} E. Joy, D. Paris, ``Spatial sampling and filtering in near field measurements,'' \textit{IEEE Trans. Antennas Propag.}, vol. 20, no. 3, pp. 253--261, 1972. 

\bibitem{planar2} J. H. Wang, ``An examination of the theory and practices of planar near-field measurement,'' \textit{IEEE Trans. Antennas Propag.}, vol. 36, pp. 746-753, 1988.


\bibitem{cylindrical1} W. Leach, D. Paris, ``Probe compensated near-field measurements on a cylinder,'' \textit{IEEE Trans. Antennas Propag.}, vol. 21, no. 4, pp. 435–445, 1973.

\bibitem{cylindrical2} J. Hansen, ``On cylindrical near-field scanning techniques,'' \textit{IEEE Trans. Antennas Propag.}, vol. 28, no. 2, pp. 231--234, Mar. 1980.


\bibitem{spherical1} F. H. Larsen, ``Probe correction of spherical near-field measurements,'' \textit{Electron. Lett.}, vol. 3, no. 14, pp. 393-395, July 1977. 

\bibitem{spherical2} J. Hansen. Spherical Near-Field Antenna Measurements; IEE Electromagnetic Wave Series 26; IET: Exeter, UK, 1988.

\bibitem{spherical3} 
O. M. Bucci, C. Gennarelli, and C. Saverese, ``Optimal Interpolation of radiated fields over a sphere,'' \textit{IEEE Trans. Antennas Propag.}, vol. 39, no. 11, pp. 1633–1643, Nov. 1991.

\bibitem{spherical4} 
A. Bangun, C. Culotta-López, ``Optimizing Sensing Matrices for Spherical Near-Field Antenna Measurements,'' \textit{IEEE Trans. Antennas Propag.}, vol. 71, no. 2, pp. 1716-1724, Feb. 2023.


\bibitem{Barakat} G. Newsam, R. Barakat, ``Essential dimension as a well-defined number of degrees of freedom of finite-convolution operators appearing in optics,'' \textit{J. Opt. Soc. Am. A}, vol. 2, no. 11, pp. 2040–2045, 1985. 

\bibitem{Miller} R. Piestun, D.A. Miller, ``Electromagnetic degrees of freedom of an optical system,'' \textit{J. Opt. Soc. Am. A} vol. 17, pp. 892–902, 2000.

\bibitem{NDF1} O. M. Bucci, G. Franceschetti, “On the degrees of freedom
of scattered fields,” \textit{IEEE Trans. Antennas Propag.}, vol. 37, no. 7, pp. 918–926, Jul. 1989.

\bibitem{NDF2} R. Pierri, R. Moretta, ``NDF of the near-zone field on a line perpendicular to the source,” \textit{IEEE Access}, vol. 9, pp. 91649–91660, 2021. 



\bibitem{local1} O.M. Bucci, C. Gennarelli, C. Savarese, ``Representation of Electromagnetic Fields over Arbitrary Surfaces by a Finite and Nonredundant Number of Samples,'' \textit{IEEE Trans. Antennas Propagat.}, vol. 46, no. 3, pp. 351--359, 1998. 

\bibitem{local2} M.D. Migliore, ``Near field antenna measurement sampling strategies: from linear to nonlinear interpolation,'' \textit{Electronics}, vol. 7, no. 10, art. no. 257, 2018. 


\bibitem{adaptive1}  D.J. Van Rensburg,  D. McNamara, G. Parsons, ``Adaptive Acquisition Techniques for Near-Field Antenna Measurements,''  in Proceedings of the 33rd Annual Antenna Measurement Techniques Association Symposium, Denver, CO, USA, 20 October 2011.

\bibitem{adaptive2} M.A. Qureshi, C.H. Schmidt, T.F. Eibert, ``Adaptive Sampling in Spherical and Cylindrical Near-Field Antenna Measurements,'' \textit{IEEE Antennas Propag. Mag.}, vol. 55, no. 1, pp. 243--249, 2013.

\bibitem{adaptive3} R. R. Alavi, R. Mirzavand, A. Kiaee, and P. Mousavi, “An adaptive data acquisition technique to enhance the speed of near-field antenna
measurement,” \textit{IEEE Trans. Antennas Propag.}, vol. 70, no. 7, pp. 5873-5883, July 2022.


\bibitem{selection1} S. Joshi, S. Boyd, ``Sensor selection via convex optimization,'' \textit{IEEE Trans. Signal Process.},  vol. 57, no. 2, pp. 451--462, 2008. 

\bibitem{selection2} A. Capozzoli, C. Curcio, A. Liseno, P. Vinetti, ``Field sampling and field reconstruction: a new perspective,'' \textit{Radio Science}, vol. 45, no. 6, pp. 131, 2010. 

\bibitem{selection3} J. Ranieri, A. Chebira, M. Vetterli, ``Near-optimal sensor placement for linear inverse problems,'' \textit{IEEE Trans. Signal Process.}, vol. 62, no. 5, pp. 1135--1146, 2014.

\bibitem{selection4} C. Jiang, Y. Soh, H. Li,  ``Sensor placement by maximal projection on minimum eigenspace for linear inverse problems,'' \textit{IEEE Trans. Signal Process.}, vol. 64, no. 21, pp. 5595--5610, 2016. 

\bibitem{selection5} J. Wang, A. Yarovoy, ``Sampling design of synthetic volume arrays for three-dimensional microwave imaging,'' \textit{IEEE Trans. Comp. Imag.}, vol. {4}, no. 4, pp. 648--660, 2018.


\bibitem{sparse1} B. Fuchs, L. L. Coq, S. Rondineau, and M. D. Migliore, “Fast
antenna far-field characterization via sparse spherical harmonic expansion,”
\textit{IEEE Trans. Antennas Propag.}, vol. 65, no. 10, pp. 5503–5510, Oct. 2017.

\bibitem{sparse2} B. Hofmann, O. Neitz, and T. Eibert, “On the minimum number
of samples for sparse recovery in spherical antenna near-field
measurements,” \textit{IEEE Trans. Antennas Propag.}, vol. 67, no. 12,
pp. 7597–7610, 2019.

\bibitem{sparse3} M. Salucci, M. D. Migliore, P. Rocca, A. Polo, and A. Massa,
“Reliable antenna measurements in a near-field cylindrical setup with a
sparsity promoting approach,” \textit{IEEE Trans. Antennas Propag.}, vol. 68,
no. 5, pp. 4143–4148, May 2020.


\bibitem{other1} L. J. Foged, F. Saccardi, F. Mioc, and P. O. Iversen, “Spherical
near field offset measurements using downsampled acquisition and
advanced NF/FF transformation algorithm,” \textit{Proc. IEEE EuCAP}, Davos, Switzerland, 2016, pp. 1–3.


\bibitem{svd2} G. Leone, R. Moretta, R. Pierri, \textquotedblleft Dimension and sampling of the near-field and its intensity over curves, \textquotedblright \textit{IEEE Open Journal of Antennas and Propagation}, vol. 3, pp. 412-424, 2022. 

\bibitem{svd3} G. Leone, F. Munno, R. Solimene and R. Pierri, \textquotedblleft A PSF Approach to Far Field Discretization for Conformal Sources," \textit{IEEE Access}, vol. 10, pp. 23394-23407, 2022.

\bibitem{MDPI2020} M.A. Maisto, R. Pierri, R. Solimene, \textquotedblleft Near-field warping sampling scheme for broad-side antenna characterization,\textquotedblright\ \ \textit{Electronics }, vol. 9, no. 1047, 2020.


\bibitem{Khare} K. Khare, N. George, Sampling-theory approach to eigenwavefronts of imaging systems. \textit{J. Opt. Soc. Am. A}, vol. 22, pp. 434–438, 2005. 

\bibitem{NDF} M.A. Maisto, R. Pierri,  R. Solimene, \textquotedblleft Near-field transverse resolution in planar source reconstructions," IEEE Transactions on Antennas and Propagation, vol. 69, no. 8, pp. 4836-4845, 2021.



\bibitem{Access2021}
M. A. Maisto, G. Leone, A. Brancaccio, R. Solimene, \textquotedblleft Efficient planar near-field measurements for radiation pattern evaluation by a warping strategy,\textquotedblright  \textit{IEEE Access}, vol. 9, pp. 62255-62265, 2021.
\end{thebibliography}
\end{document}